# Magnetic Seismology of Interstellar Gas Clouds: Unveiling a Hidden Dimension

Aris Tritsis,[1,2,3*] Konstantinos Tassis[1,3*]

[1]Department of Physics and Institute of Theoretical and Computational Physics,
University of Crete, PO Box 2208, 71003 Heraklion, Crete, Greece.
[2]Research School of Astronomy and Astrophysics,
Australian National University, Canberra, ACT 2611, Australia.
[3]Institute of Electronic Structure and Laser and Institute of Astrophysics,
Foundation for Research and Technology-Hellas, PO Box 1527, 71110 Heraklion, Crete, Greece.

[*]To whom correspondence should be addressed; E-mail: tassis@physics.uoc.gr, Aris.Tritsis@anu.edu.au

**Stars and planets are formed inside dense interstellar molecular clouds, by processes imprinted on the 3-dimensional (3D) morphology of the clouds. Determining the 3D structure of interstellar clouds remains challenging, due to projection effects and difficulties measuring their extent along the line of sight. We report the detection of normal vibrational modes in the isolated interstellar cloud Musca, allowing determination of the 3D physical dimensions of the cloud. Musca is found to be vibrating globally, with the characteristic modes of a sheet viewed edge-on, not a filament as previously supposed. We reconstruct the physical properties of Musca through 3D magnetohydrodynamic simulations, reproducing the observed normal modes and confirming a sheet-like morphology.**

Astronomical objects are seen in two-dimensional projection on the plane of the sky. This is particularly problematic for studies of the interstellar medium, because the 3-dimensional (3D) structure of interstellar clouds encodes information regarding the physical processes (such as magnetic forces, turbulence and gravity) that dominate the formation of stars and planets. We seek a solution to this problem by searching for resonant magnetohydrodynamic (MHD) vibrations in an isolated interstellar cloud and by analysing its normal modes. Normal modes have been used extensively to describe and analyse various systems in the physical sciences, from quantum mechanics and helioseismology, to geophysics and structural biology. Normal modes have been observed in the interstellar medium (ISM) in two small pulsating condensations (Bok globules) located inside two molecular clouds (i.e. dense enough interstellar clouds allowing the formation of molecular hydrogen) (*1, 2*). Further applications have been limited, because molecular clouds usually exhibit a complex morphology including filamentary structures, as a result of turbulent mixing and shock interaction (*3, 4*).

Recent wide-field radio observations of molecular clouds (*5*) have unveiled the presence of well-ordered, quasi-periodically spaced elongations, termed striations, on the outskirts of clouds. The thermal dust continuum emission survey of nearby molecular clouds by the *Herschel* Space Observatory has shown that striations are a common feature of clouds (*6-10*), often associated with denser filaments (*7-11*), inside which stars are formed. Complementary polarimetric studies have revealed that striations are always well-aligned with the cloud's magnetic field projected onto the plane of the sky (*5,7-12*).

From a theoretical perspective, the only viable mechanism for the formation of striations involves the excitation of fast magnetosonic waves (i.e. longitudinal magnetic pressure waves) (*13*). Compressible fast magnetosonic waves can be excited by non-linear coupling with Alfvén waves (i.e. incompressible transverse waves along magnetic field lines) and/or perturbations created by self-gravity in an inhomogeneous medium. These magnetosonic waves compress the gas and form ordered structures parallel to magnetic field lines, in agreement with observations of striations (*5,7-12*).

Once magnetosonic waves are excited, they can be reflected in regions of varying Alfvén speed (defined as $v_A = B/\sqrt{4\pi\rho}$ where B is the magnetic field and $\rho$ is the density of the medium), setting up normal modes, just like vibrations in a resonating chamber. In regions where striations appear to be unassociated with denser structures (e.g. H I clouds), this resonating chamber may be the result of external pressure confinement by a more diffuse, warmer medium. However, boundaries can also be naturally created in the case of a contracting self-gravitating cloud as a result of steep changes in density and magnetic field, that in turn lead to sharp variations in the velocity of propagation of these waves (*14*). Any compressible fast magnetosonic waves excited during the formation of the cloud will then be trapped, thus resulting in striations in the vicinity of denser structures.

Fast magnetosonic waves travelling in both directions perpendicular to the magnetic field are coupled (*13*). By considering a rectangular box the spatial frequencies of normal modes ($k_{nm}$) are given by:

$$k_{nm} = \sqrt{\left(\frac{\pi n}{L_x}\right)^2 + \left(\frac{\pi m}{L_y}\right)^2} \quad (1)$$

where the ordered component of the magnetic field is considered to be along the $z$ direction, and $L_x$ and $L_y$ are the sizes of the box in the $x$ and $y$ directions respectively, whilst $n$ and $m$ are integers ranging from zero to infinity. By considering a rotation matrix it can be shown that the spatial frequencies seen in the power spectra of cuts perpendicular to the long axis of striations are independent of the orientation of the cloud (*14*).

We analyse these magnetohydrodynamic striations seen in Musca (G301.70-7.16), a molecular cloud located ~ 150 to 200 pc from Earth (*15, 16*). Due to its elongated and ordered morphology, and its low column density (i.e. integrated volume density along the line of sight), Musca is considered to be the prototype of a filamentary/cylindrical molecular cloud (*9,17-20*) which is used as a comparison by many theoretical models. Musca has been mapped by *Herschel* as part of the Gould Belt Survey (*9*) and exhibits clear striations oriented perpendicularly to the main body of the cloud. We have re-analysed the archival data (*9*); `Fig. 1` shows the *Herschel*-Spectral and Photometric Imaging Receiver 250 μm dust emission map of Musca. We have considered cuts perpendicular to the long axis of striations inside the green rectangle in `Fig. 1` in order to study their spatial power spectra. We have verified that our selection does not introduce biases by considering cuts perpendicular to the long axis of the striations and studying their spatial power spectra in other regions as well (*14*).

The normalized power spectra from each cut and the distribution of the identified peaks are shown in `Fig. 2A` and `Fig. 2B` respectively. From Eq. (1) and assuming $L_x$ is the largest dimension of the cloud, the smallest possible wavenumber is obtained for $(n, m) = (1, 0)$. Thus, the first peak in `Fig. 2B` has to correspond to $(n, m) = (1, 0)$ yielding $L_x = 8.2 \pm 0.3$ pc. This value is consistent with the observed size of the cloud on the plane of the sky, which is variously reported to be from 6.5 to 7.85 pc when scaled to our adopted cloud distance of 150 pc (*9,18,21*). The second peak could correspond to either $(n, m) = (0, 1)$ or, in the case of a cylindrical cloud with $L_x \gg L_y$ (and thus $k_{01} \gg 1$), to $(n, m) = (2, 0)$. However, with $L_x \sim 8$ pc, the $(n, m) = (2, 0)$ peak is expected at $k \sim 0.8$ (pc)$^{-1}$, much higher than the actual location of the second peak. Thus, this second peak has to correspond to $(n, m) = (0, 1)$. Inserting $(n, m) = (0, 1)$ and the value of the second peak in Eq. (1), the hidden, line of sight dimension $L_y$ is deduced as $6.2 \pm 0.2$ pc, comparable to the largest dimension of the cloud. The other normal modes with their uncertainties determined through error propagation are predicted analytically by inserting these values for $L_x$ and $L_y$ into Eq. (1) and are over-plotted in `Fig. 2B`. Therefore, Musca, previously considered to be a prototypical cylindrical/filamentary cloud, is instead a sheet-like structure seen edge-on.

In `Fig. 2B` we plot all the normal modes up to $(n, m) = (2, 2)$. However, we find good agreement between the predicted wavenumbers and observations up to the first few modes with $n$ or $m = 4$ corresponding to physical scales of the order of 1.6 pc (see Fig. S1). However, the shape of the cloud is more complicated than an idealized rectangle, exhibiting higher order structure on smaller scales, so the normal modes may be better modelled using a rectangle with rounded edges or an ellipse. Thus, Eq. (1) is an approximation that applies only to the normal modes with small spatial frequencies, i.e. large physical scales. At spatial frequencies higher than $\sim 2$ (pc)$^{-1}$ the density of normal modes becomes so high that they cannot be identified in either the observations or the theoretical predictions (the uncertainties overlap for all predicted modes).

Through ideal (i.e. non-dissipative) MHD simulations including self-gravity (*14*), we have constructed a 3D model of Musca, including the dense structure and striations in the low-density parts. In `Fig. 3` we show the column density map from our simulation, which reproduces the observed dimensions of the cloud. A 3D representation of the volume density of the model of Musca is shown in `Fig. 4`. As intuitively expected from the normal mode analysis of the observations, the shape of the cloud is that of a rectangle with rounded edges.

The maximum column density in the simulation, assuming an edge-on view, is $1.9 \times 10^{22}$ cm$^{-2}$. For comparison, the maximum column density derived observationally from the dust emission maps (*9*) is $\sim 1.6 \times 10^{22}$ cm$^{-2}$. The maximum volume number density in the simulation is $\sim 2 \times 10^{3}$ cm$^{-3}$, high enough for molecules to be collisionally excited and therefore observed using their rotational emission lines. Molecular line observations of the Musca molecular cloud are limited to CO, including several isotopologues, and $NH_3$ (*17-20*); the latter is only observed towards the densest core of Musca. The number densities required to excite CO and $NH_3$ lines are $\sim 10^{2}$ and $10^{3}$ cm$^{-3}$ respectively (*22*) which are easily reached in our simulated model of the cloud. In contrast to the sheet-like structure, in order to reproduce the observed column density in any filament-like geometry, the number density has to be of the order of $5 \times 10^{4}$ cm$^{-3}$ or higher (*18*). This value is well above the density threshold for star formation for clouds in the Gould - Belt and a density threshold derived specifically for Musca (*23*). More evident star formation activity should be observed if Musca was a filament. Moreover, if the 3D shape of Musca was that of a filament, $NH_3$ would be easily excited and observed throughout the ridge of the dense structure.

We use a suite of simulations of clouds of different shapes to validate our analysis and verify that Eq. (1) can be used to extract the correct cloud dimensions (*14*). In each of our simulations, the known dimensions of the clouds are recovered by the simulated normal-mode analysis. In contrast to the distribution of peaks seen in `Fig. 2B`, in cylindrical clouds ($L_y \ll L_x$) the first few peaks at low spatial frequencies are all multiples of the first peak. The first few peaks for cylindrical clouds are only due to the largest dimension of the cloud, resulting in a sparser distribution of peaks than the sheet geometry (see Fig. S4). This is both quantitatively and qualitatively different to the distribution seen in the Musca data (see `Fig. 2`), strengthening the case that the intrinsic shape of Musca is sheet-like. Sheet-like structures are common in turbulent clouds as they may represent planar-like shocks from processes such as supernova explosions or expanding ionization regions, or simply result from accretion along magnetic field lines (*4, 24, 25*).

For decades, the determination of the 3D shape of clouds has been pursued through statistical studies (*26-28*), which do not provide information on a cloud by cloud basis. Other proposed methods (*29, 30*) rely on complex chemical and/or radiative processes and thus depend on numerous assumptions. With its 3D geometry now determined, Musca can be used to test theoretical models of interstellar clouds.

### References and Notes

1. C. J. Lada, E. A. Bergin, J. F. Alves, T. L. Huard, *Astrophys. J.*, **586**, 286-295 (2003)

2. E. D. Aguti, C. J. Lada, E. A. Bergin, J. F. Alves, & M. Birkinshaw, *Astrophys. J.*, **665**, 457-465 (2007)

3. P. André, A. Men'shchikov, S. Bontemps, et al. *Astron. Astrophys.*, **518**, L102 (2010)

4. C. Federrath, *Mon. Not. R. Astron. Soc.*, **457**, 375-388 (2016)

5. P. F. Goldsmith, M. Heyer, G. Narayanan, et al., *Astrophys. J.*, **680**, 428-445 (2008)

6. M.-A. Miville-Deschênes, P. G. Martin, A. Abergel, et al., *Astron. Astrophys.*, **518**, L104 (2010)

7. C. Alves de Oliveira, N. Schneider, B. Merín et al., *Astron. Astrophys.*, **568**, A98 (2014)

8. P. Palmeirim, P. André, J. Kirk et al., *Astron. Astrophys.*, **550**, A38 (2013)

9. N. L. J. Cox, D. Arzoumanian, P. André, et al., *Astron. Astrophys.*, **590**, A110 (2016)

10. J. Malinen, L. Montier, J. Montillaud et al., *Mon. Not. R. Astron. Soc.*, **460**, 1934-1945 (2016)

11. N. Schneider, T. Csengeri, M. Hennemann, et al., *Astron. Astrophys.*, **540**, L11 (2012)

12. G. V. Panopoulou, I. Psaradaki, K. Tassis, *Mon. Not. R. Astron. Soc.*, **462**, 1517-1529 (2016)

13. A. Tritsis, K. Tassis, *Mon. Not. R. Astron. Soc.*, **462**, 3602-3615 (2016)

14. Materials and methods are available as supplementary materials on Science Online.

15. J. Knude, E. Hog, *Astron. Astrophys.*, **338**, 897-904 (1998)

16. J. C. Gregorio Hetem, G. C. Sanzovo, J. R. D. Lepine, *Astron. Astrophys.*, **76** (suppl.), 347-363 (1988)

17. E. M. Arnal, R. Morras, J. R. Rizzo, *Mon. Not. R. Astron. Soc.*, **265**, 1 (1993)

18. J. Kainulainen, A. Hacar, J. Alves et al., *Astron. Astrophys.*, **586**, A27 (2016)

19. A. Hacar, J. Kainulainen, M. Tafalla, H. Beuther, J. Alves, *Astron. Astrophys.*, **587**, A97 (2016)

20. D. A. Machaieie, J. W. Vilas-Boas, C. A. Wuensche et al., *Astrophys. J.*, **836**, 19 (2017)

21. Planck Collaboration, P. A. R. Ade, N. Aghanim et al., *Astron. Astrophys.*, **586**, A136 (2016)

22. Y. L. Shirley, *Publications of the Astronomical Society of the Pacific*, **127**, 299 (2015)

23. J. Kainulainen, C. Federrath, & T. Henning, Science, **344**, 183-185 (2014)

24. F. Nakamura, & Z.-Y. Li, *Astrophys. J.*, **687**, 354-375 (2008)

25. T. C. Mouschovias, *Astrophys. J.*, **373**, 169-186 (1991)

26. P. C. Myers, G. A. Fuller, A. A. Goodman, P. J. Benson, *Astrophys. J.*, **376**, 561-572 (1991)

27. C. E. Jones, S. Basu, J. Dubinski, *Astrophys. J.*, **551**, 387-393 (2001)

28. K. Tassis, C. D. Dowell, R. H. Hildebrand, L. Kirby, J. E. Vaillancourt, *Mon. Not. R. Astron. Soc.*, **399**, 1681-1693 (2009)

29. J. Steinacker, A. Bacmann, T. Henning, R. Klessen, M. Stickel, *Astron. Astrophys.*, **434**, 167-180 (2005)

30. D. Li, P. F. Goldsmith, *Astrophys. J.*, **756**, 12 (2012)

31. M. J. Griffin, A. Abergel, A. Abreu, et al., *Astron. Astrophys.*, **518**, L3 (2010)

32. J. D. Scargle, *Astrophys. J.*, **263**, 835-853 (1982)

33. R. H. D. Townsend, *Astrophys. J. Suppl. Ser.*, **191**, 247-253 (2010)

34. B. Fryxell, K. Olson, P. Ricker, et al., *Astrophys. J. Suppl. Ser.*, **131**, 273-334 (2000)

35. A. Dubey, R. Fisher, C. Graziani, et al., *Numerical Modeling of Space Plasma Flows*, **385**, 145 (2008)

36. D. Lee, *Journal of Computational Physics*, **243**, 269-292 (2013)

37. P. L. Roe, *Journal of Computational Physics*, **43**, 357-372 (1981)

38. B. van Leer, *Journal of Computational Physics*, **32**, 101 (1979)

39. M. J. Turk, B. D. Smith, J. S. Oishi, et al., *Astrophys. J. Suppl. Ser.*, **192**, 9 (2011)

40. P. Ramachandran, G. Varoquaux, *Astrophysics Source Code Library*, 1205.008 (2012)

41. S. Chandrasekhar, E. Fermi, *Astrophys. J.*, **118**, 113 (1953)

42. R. H. Hildebrand, L. Kirby, J. L. Dotson, M. Houde, J. E. I. Vaillancourt, *Astrophys. J.*, **696**, 567-573 (2009)

43. Planck Collaboration, P. A. R. Ade, N. Aghanim, et al., *Astron. Astrophys.*, **586**, A138 (2016)

# Acknowledgments

We thank V. Pavlidou, G. Panopoulou, V. Charmandaris, N. Kylafis, A. Zezas, E. Economou, J. Andrews, S. Williams, P. Sell, D. Blinov, I. Liodakis, T. Mouschovias and the three anonymous referees for comments that helped improve this paper. Usage of the Metropolis HPC Facility at the Crete Center for Quantum Complexity and Nanotechnology of the University of Crete, supported by the European Union Seventh Framework Programme (FP7-REGPOT-2012-2013-1) under grant agreement no. 316165, is acknowledged.

# Funding

K.T. and A.T. acknowledge support by the 7$^{th}$ Framework Programme through Marie Curie Career Integration Grant PCIG- GA-2011-293531 “Onset of Star Formation: Connecting Theory and Observations". A.T. acknowledges funding from the European Research Council under the European Union’s Seventh Framework Programme (FP/2007-2013)/ERC Grant Agreement n. 617001.

# Author Contributions

A.T. performed the numerical simulations, the analysis of the observations and wrote the text. K.T. contributed with the interpretation of the results and the writing of the text.

# Competing interests

The authors declare no conflicts of interest.

# Data and materials availability

All observational data used in this research are from the Herschel Gould Belt Survey project, publicly available at http://archives.esac.esa.int/hsa/whsa/#home (observation ID: 1342216012). All simulation outputs and setup files are available at https://doi.org/10.6084/m9.figshare.5950360. The Flash software used in this work was in part developed by the Department of Energy's National Nuclear Security Administration - Advanced Simulation and Computing Program; Office of Advanced Scientific Computing Research Flash Center at the University of Chicago and was obtained from http://flash.uchicago.edu/site/. The "YT" analysis toolkit is available at http://yt-project.org/#getyt and "MAYAVI2" is available at https://github.com/enthought/mayavi.

# Supplementary Materials

www.sciencemag.org

Materials and Methods

Figs. S1, S2, S3, S4

References (31-43) [Note: The numbers refer to any additional references cited only within the Supplementary Materials]

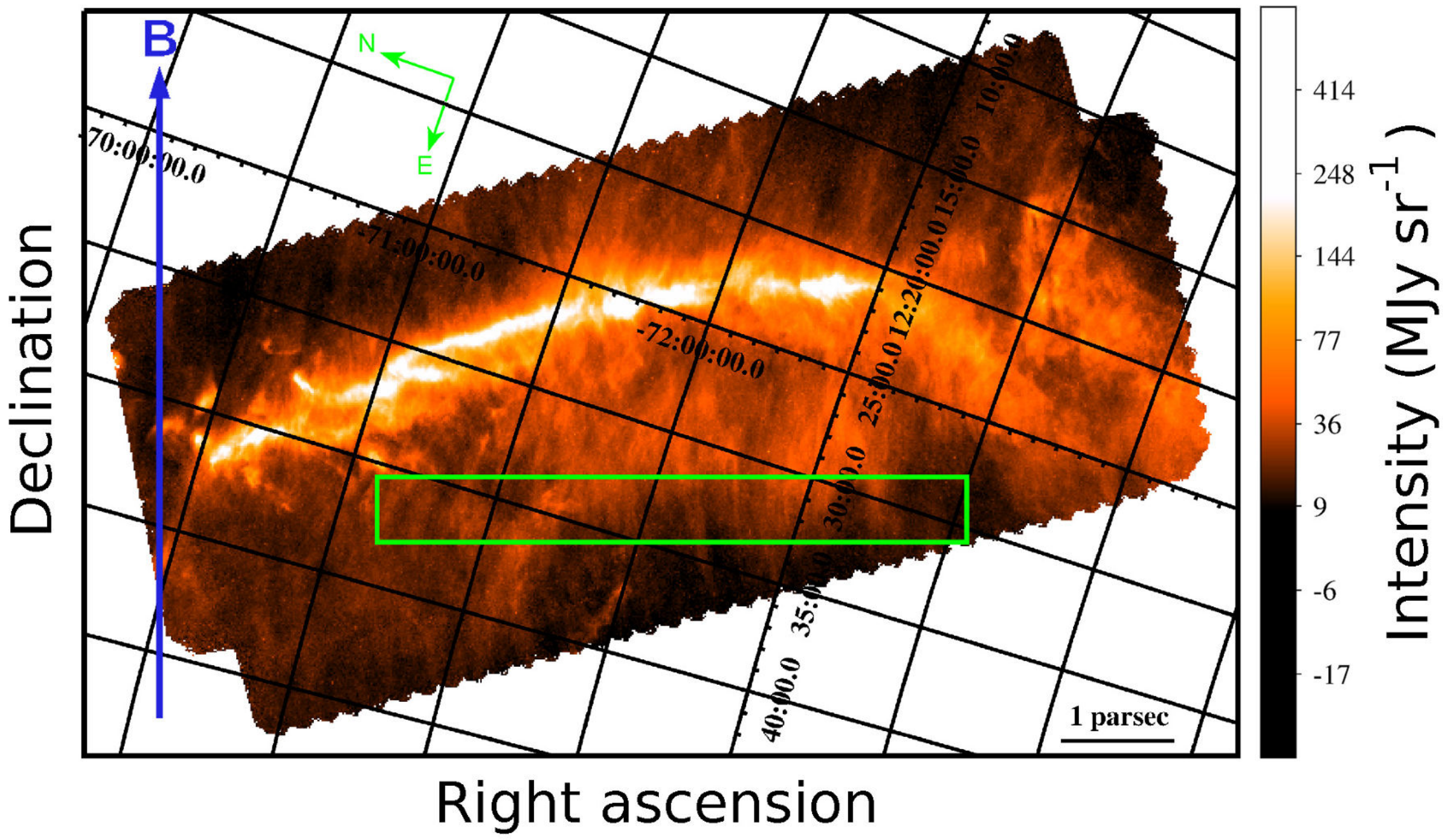

**Fig. 1. The Musca molecular cloud.** *Herschel* 250 µm dust emission map of the Musca molecular cloud (in mega Jansky per steradian) showing both striations and the dense elongated structure. The green rectangle marks the region where we have performed our normal-mode analysis, the blue arrow shows the mean direction of the magnetic field projected onto the plane of the sky (*9*). Grid lines show equatorial coordinates.

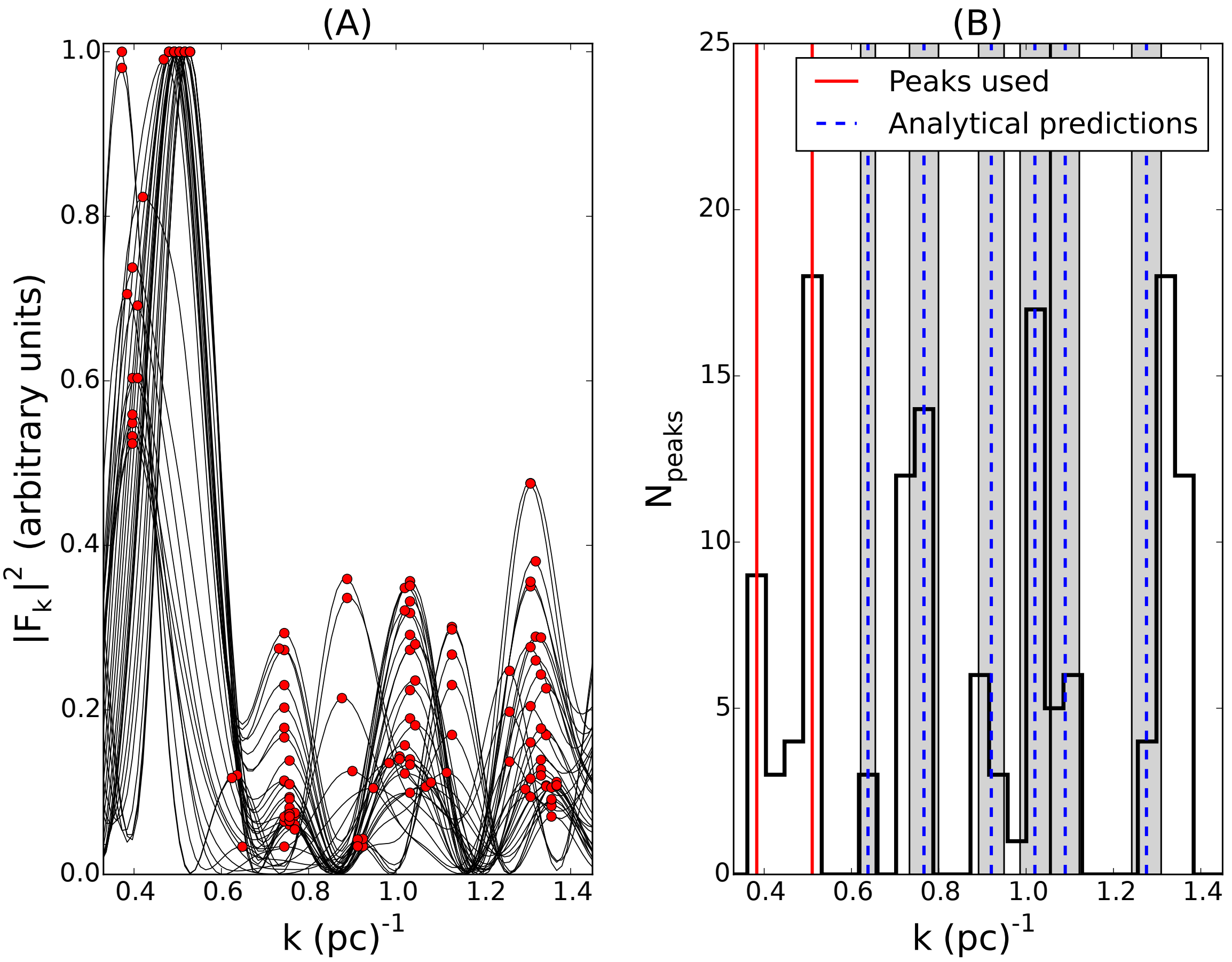

**Fig. 2. Comparison of observed normal modes with analytical solution.** (A) Normalized power spectra (black lines) of cuts trough the observations perpendicular to the striations; peaks we identified are marked with red dots. (B) Distribution of peaks at different spatial frequencies. The red lines depict the values used to derive the dimensions of the cloud. The blue dashed lines show the rest of the normal modes [up to $(n, m) = (2, 0)$], predicted analytically from Eq. (1) given the cloud dimensions derived from the first two peaks. Shaded regions indicate the $1\sigma$ regions of the analytical predictions due to uncertainties in the determination of the location of the first two peaks, propagated through Eq. (1). The bin size is comparable to the standard deviations of the points comprising the first two peaks.

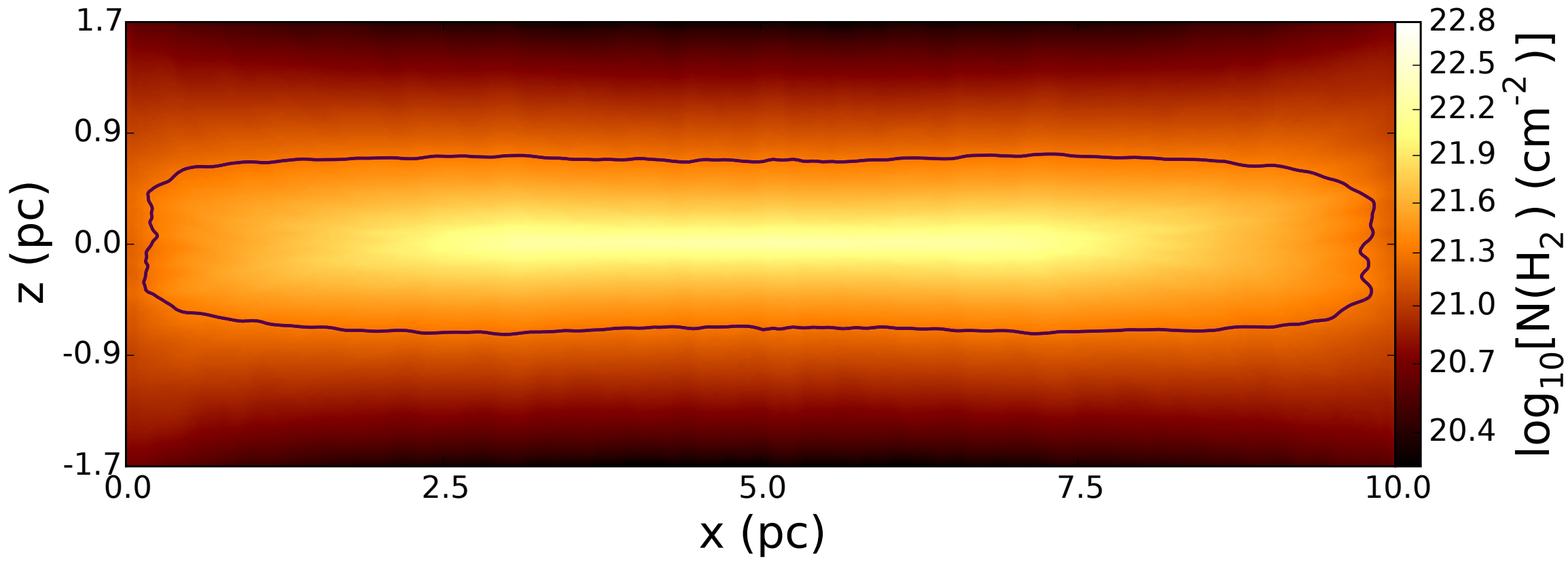

**Fig. 3. Column density map of the model of Musca molecular cloud.** Edge-on view of the molecular gas column density from our MHD simulation of a sheet-like structure. The color bar shows the logarithm of the column density. The purple contour marks the region with $N(H_2) > 2 \times 10^{21}$ $cm^{-2}$ used to identify Musca main, dense filament (*9*). The magnetic field is along the z axis and the time of the snapshot since the beginning of the simulation is $\sim 2.7$ Myrs.

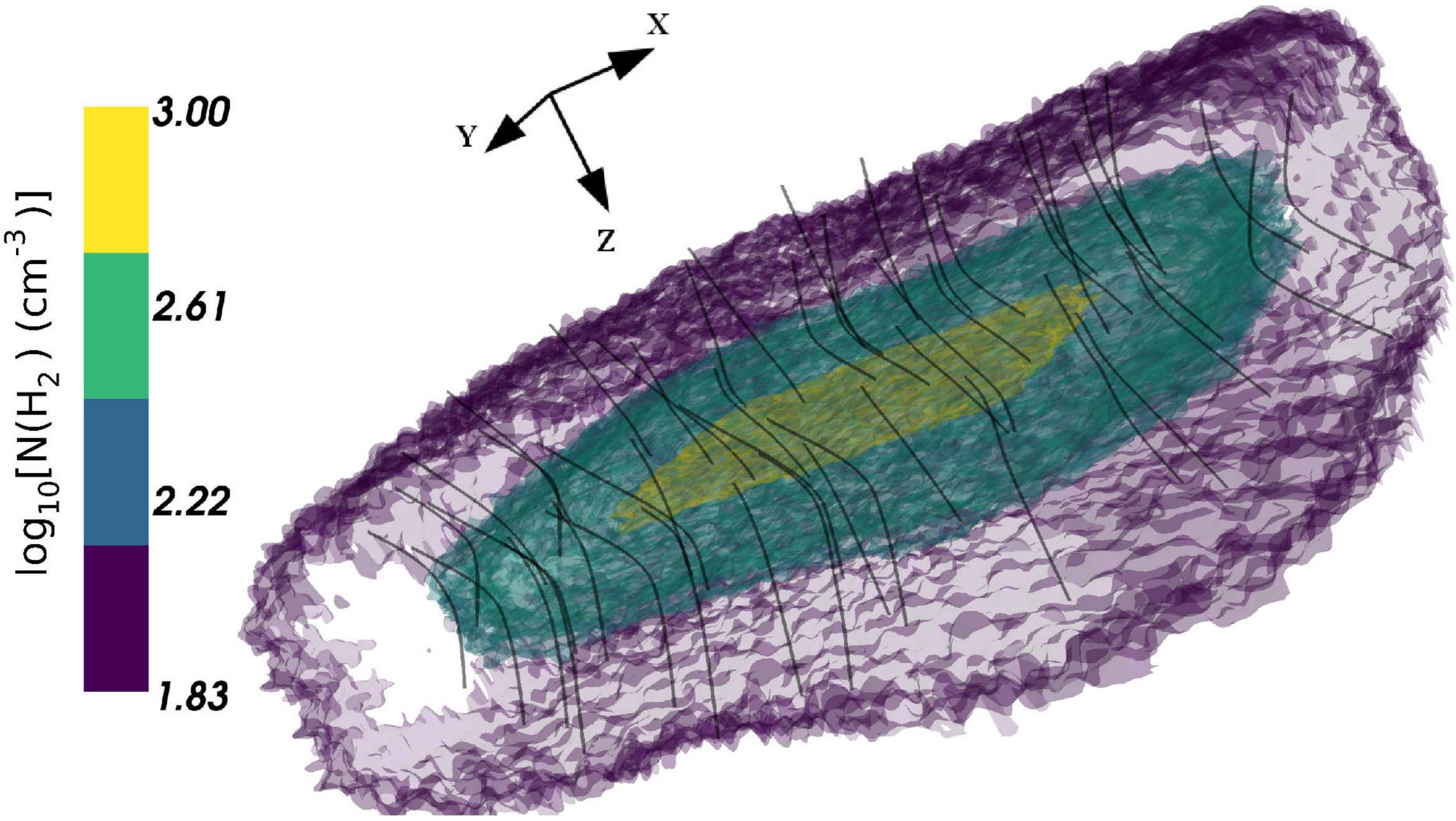

**Fig. 4. 3D model of Musca molecular cloud.** Logarithmic 3D volume density in our MHD simulation of the Musca cloud. Density isosurfaces are set at 90%, 75%, 70% and 55% of the logarithm of the maximum number density. Black lines represent the magnetic field.

# Supplementary Materials for

## Magnetic Seismology of Interstellar Gas Clouds: Unveiling a Hidden Dimension

Aris Tritsis, Konstantinos Tassis

correspondence to: tassis@physics.uoc.gr, Aris.Tritsis@anu.edu.au

**This PDF file includes:**

## Materials and Methods

Observations

We use Level 3 *Herschel*-Spectral and Photometric Imaging Receiver (*31*) 250 μm dust emission data, obtained directly from the *Herschel* Science Archive. Observations were carried out on 2011 March 13 and 14, at a scanning speed of 30 arcsec $s^{-1}$. The spatial resolution of *Herschel* at 250 μm is 18 arcsec. We define a Cartesian coordinate system where the $z$ axis is aligned with the long axis of the striations and thus the plane-of-the-sky (POS) magnetic field, as probed by polarization measurements (*9*), the $y$ axis is parallel to the line-of-sight (LOS), and the $x$ axis is perpendicular to the striations on the POS.

For deriving the normal modes we select a region south-east of the main filamentary structure (Fig. 1) and consider cuts along the $x$ axis. Gravity and other effects such as shocks resulting from the formation of the cloud will impact the structure of regions adjacent to the dense filament. For this reason, the region has been selected based on three criteria: *a*) it is as far away from the dense filamentary structure as possible so that additional forces can be ignored compared to the effect of MHD waves *b*) it has the largest possible length so that we can retrieve all spatial frequencies and *c*) the cuts are perpendicular to striations throughout, so that small variations in the orientation angle of striations do not affect our results. In order to increase the signal-to-noise ratio in our analysis we average the intensity of $z$ pixel values in every three adjacent cuts along the $x$ direction. We then compute the spatial power spectrum of each averaged cut using the Lomb-Scargle periodogram technique (*32, 33*).

We have verified that increasing the size of the region, selecting a region north of the main filamentary structure and the uncertainties in the orientation of striations result in only small variations in the derived dimensions. In Fig. S1 we show the normalized power spectra (panel A) and a histogram of the identified peaks (panel B) from a region north-west of the main dense structure. The range of the $x$-axis in the right panel was increased in order to show the agreement between the theoretical predictions and observations for smaller physical scales. In this region, due to the curvature of Musca, there is an uncertainty with regard to the orientation angle of striations. In order to minimize this effect we also considered cuts at angles (within a range of 6°) other than that where the orientation angle of striations peaks. All the cuts had the same length and the final distribution shown was constructed from all cuts. The derived dimensions remain the same within uncertainties ($L_x = 8.4 \pm 0.3$ and $L_y = 6.4 \pm 0.4$ pc).

The uncertainty in the derived dimensions and in the analytically predicted normal modes is calculated from the uncertainty in the spatial frequencies of the first two peaks through error propagation. For the uncertainty of these spatial frequencies we take the standard deviations of the points comprising each of the first two peaks. The adopted distance of Musca in our analysis is 150 pc (*15*). Earlier estimates placed Musca at 200 pc (*16*). Adopting this larger estimate has no effect on the quality of the fitting of normal modes from Eq. (1) since this only depends on the relative distance between peaks. However, because angular size translates to a different physical scale the derived dimensions change to $L_x = 10.4 \pm 0.4$ and $L_y = 8.3 \pm 0.2$ pc.

More disk-like rather than triaxial oblate clouds should exhibit normal modes given by:

$$k_{nm} = \frac{j_{nm}}{R} \quad \text{(S1)}$$

where $R$ is a mean radius and $j_{nm}$ are the roots of the derivative of the Bessel function of first kind. In the case of Musca, unlike Eq. (1), Eq. (S1) fails to fit the observed modes, thus suggesting a more triaxial shape.

Regarding the projection angle, a rotation of the x-y coordinate system around the z-axis will leave the observed spatial frequencies unchanged:

$$\begin{bmatrix} k_x' \\ k_y' \end{bmatrix} = \begin{bmatrix} \cos(\theta) & -\sin(\theta) \\ \sin(\theta) & \cos(\theta) \end{bmatrix} \begin{bmatrix} k_x \\ k_y \end{bmatrix} \Rightarrow \begin{matrix} k_x' = k_x \cos(\theta) - k_y \sin(\theta) \\ k_y' = k_x \sin(\theta) + k_y \cos(\theta) \end{matrix} \quad \text{(S2)}$$

where $k_x$ and $k_y$ are the wavenumbers in the $x$ and $y$ directions respectively, $\theta$ is the angle between the cloud's y-axis and the LOS and $k_x'$ and $k_y'$ are the new wavenumbers in the $x$ and $y$ directions at this new angle $\theta$. Thus, the new observed spatial frequencies ($k'$) at this new angle $\theta$ will be:

$$k' = \sqrt{k_x'^2 + k_y'^2} = \sqrt{k_x^2(\sin^2(\theta) + \cos^2(\theta)) + k_y^2(\sin^2(\theta) + \cos^2(\theta))} = k \quad \text{(S3)}$$

As a result, the spatial frequencies are independent of the projection angle.

Cautionary remarks

We make two cautionary remarks regarding the method. The first is related to the orientation angle of striations. Although striations in Musca and in all clouds are well ordered, their orientation angle has a non-zero spread. When the cuts are not perpendicular to the striations we are probing the desired scales over cos($\phi$), where $\phi$ is the angle between the cut and the long axis of striations. Thus, each peak in the distribution of the spatial frequencies is also expected to have a non-zero spread. The fact that the orientation angle of striations has a non-zero spread is not an artifact of the analysis of the observations but it is to be physically expected because fast magnetosonic waves can propagate in directions other than exactly perpendicular to the magnetic field. However, these modes ultimately escape the region along field lines.

The second cautionary remark is related to the phase difference between different wave modes. For a particular cut, the phase of a wave mode can be such that the displacement is zero (or close to zero) and it cannot be retrieved in the power spectrum. This can be seen in Figures 2A, S1A and S2A where not all spectra appear to go through all peaks. In order to probe all wave modes, a sufficiently large region has to be selected. This phase difference arises from the different time scales involved for different modes.

Simulations

We have used the astrophysical code FLASH 4.4 (*34, 35*) to perform three dimensional, ideal MHD simulations with self-gravity in Cartesian coordinates. To solve the ideal MHD equations we use the unsplit staggered mesh algorithm (*36*). This scheme has advantages over truncation-error methods and $\nabla \cdot \vec{B} = 0$ is satisfied at all times to machine precision. In the interest of reducing computational cost and because we are not interested in the fragmentation of any dense structures created within the cloud we use the standard FLASH multipole algorithm to solve Poisson's equation. To account for all waves that can arise in the MHD equations we use Roe's solver for the Riemann problem (*37*). To minimize numerical diffusivity we use van Leer's flux limiter (*38*) and third order interpolation. Because Musca is an isolated cloud, we model it as such by forcing normal velocity components to zero at the boundaries. This is achieved by setting all boundary types to diode, i.e. non-reflective type of boundaries. Thus, our choice ensures that any reflection of waves occurs at boundaries naturally created due to the contraction of the cloud and not at the edges of our simulation box. Finally, diode boundary conditions ensure there is no mass influx during the evolution of our simulations. Simulations have been performed on a fixed resolution grid with 256 × 256 × 256 cells. We have verified that numerical resolution does not affect our results by performing convergence tests. For exporting simulation data from Hierarchical Data Format, version 5 we use the "YT" analysis toolkit (*39*). 3D plots were created using "MAYAVI2"(*40*).

In our simulations we again define a Cartesian coordinate system such that the direction of the ordered magnetic field is along the $z$ direction and $y$ represents the LOS dimension. Through the Chandrasekhar-Fermi method (*41*) and its more updated interpretation (*42*), the ordered value of the POS magnetic field towards the Musca molecular cloud has been observationally determined to be $12 \pm 5$ and $27 \pm 11$ μG respectively (*43*). These estimates assume a number density of 100 $cm^{-3}$. We adopt a conservative value of 7 μG as an initial condition for our simulations which is within the aforementioned observational limits. In all simulations, we additionally introduce a perturbation in the $x$ and $y$ components of the magnetic field as:

$$B_x(z)=B_y(z)=\delta B-\delta B\sin(k_z z) \tag{S4}$$

where $B_x$ and $B_y$ are the $x$ and $y$ components of the magnetic field, $k_z = \pi / L_z$ and $\delta B$ was set equal to 10% of the value of the unperturbed field. This setup implies an Alfvén wave passing through the computational box with wavelength twice the size of the $z$ direction. A constant temperature of 14 K is used in all simulations. Random perturbations with maximum amplitude of 40% of the background value are introduced in density and thermal pressure in a self-consistent manner such that the gas remains isothermal. All velocity components are initially set to zero in all simulations. Thus, the Alfvén Mach number is smaller than one at all times during the evolution of the simulations, which is the regime where striations are formed.

<u>3D model of Musca</u>

The initial number density in our simulation is set equal to 100 $cm^{-3}$. The dimensions of our simulation box are $L_x = 10.0$, $L_y = 8.0$ and $L_z = 3.5$ pc. The $x$ and $y$ dimensions of the simulation box are thus higher than the values we have derived observationally. However, because of contraction due to self-gravity and pinching of the field lines, the region where fast magnetosonic waves are trapped is smaller than the simulation box.

In Fig. S2 we show the normalized power spectra (panel A) and the histogram of the identified peaks (panel B) from striations in this 3D model of Musca, computed from the simulated column density map. The derived $L_x$ and $L_y$ dimensions are $7.7 \pm 0.1$ and $5.6 \pm 0.1$ pc respectively, in agreement with the dimensions of the resonating region as measured from abrupt changes in the propagating velocity of the waves ($L_x = 7.6$ and $L_y = 5.8$ pc; see Fig. S3). These values correspond to Fig. 3 when the elapsed time is ~ 2.7 Myrs, 0.3 Myrs after the dense structure (volume number density of $\sim 10^3$ $cm^{-3}$) has formed. Since the morphology of the dense structure is that of a rectangle with rounded edges (Fig. 4), similar to the observations, Eq. (1) applies only to large physical scales.

In Fig. S3 we show the velocity of propagation of fast magnetosonic waves in the mid-plane along the *y* direction. Changes of magnetic field strength along the *x* direction due to pinching of field lines result in variations in the propagation speed and thus define boundaries where magnetosonic waves are reflected. In 3D the boundaries resemble a square bowl with rounded edges. Fig. S3 shows the region where we performed our normal-mode analysis, at a comparable distance from the main dense structure as in the observations.

<u>Benchmarking</u>

We have performed an additional simulation considering a filament-like structure with different initial conditions for density, in order to benchmark our analysis. However, this filament-like simulation does not match Musca. The dimensions of the box of this filament simulation are $L_x = 6.0$, $L_y = 2.0$ and $L_z = 2.0$ pc and the unperturbed value of density is set to 200 $cm^{-3}$. The column density map as well as the normalized power spectra are shown in Fig. S4. All peaks are due to and can be fitted by the largest dimension of the cloud alone, since the modes from the shortest direction lie at much higher spatial frequencies ($k > 6$ $pc^{-1}$). Thus, all the peaks shown in Fig. S4B correspond to $m = 0$ in Eq. (1). The normal-mode analysis of striations in the simulation yields $L_x = 4.2 \pm 0.1$ pc and $L_y \ll L_x$, again in agreement with the size of the resonating region.

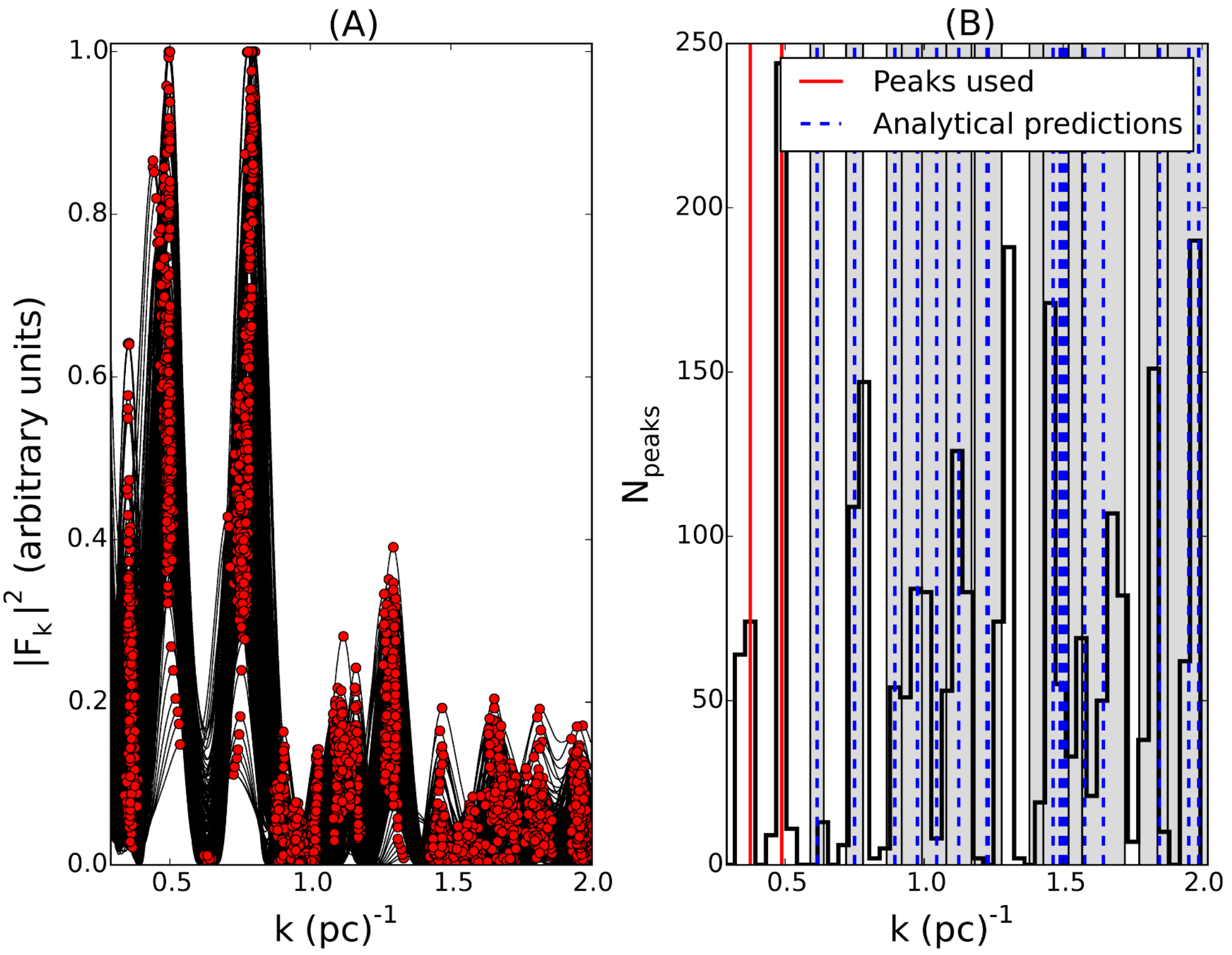

**Fig. S1.**

Normal modes from a region of striations above the dense cloud. The lines in the left and right panels are the same as Fig. 2. The agreement between theoretical predictions and observations continues in smaller physical scales.

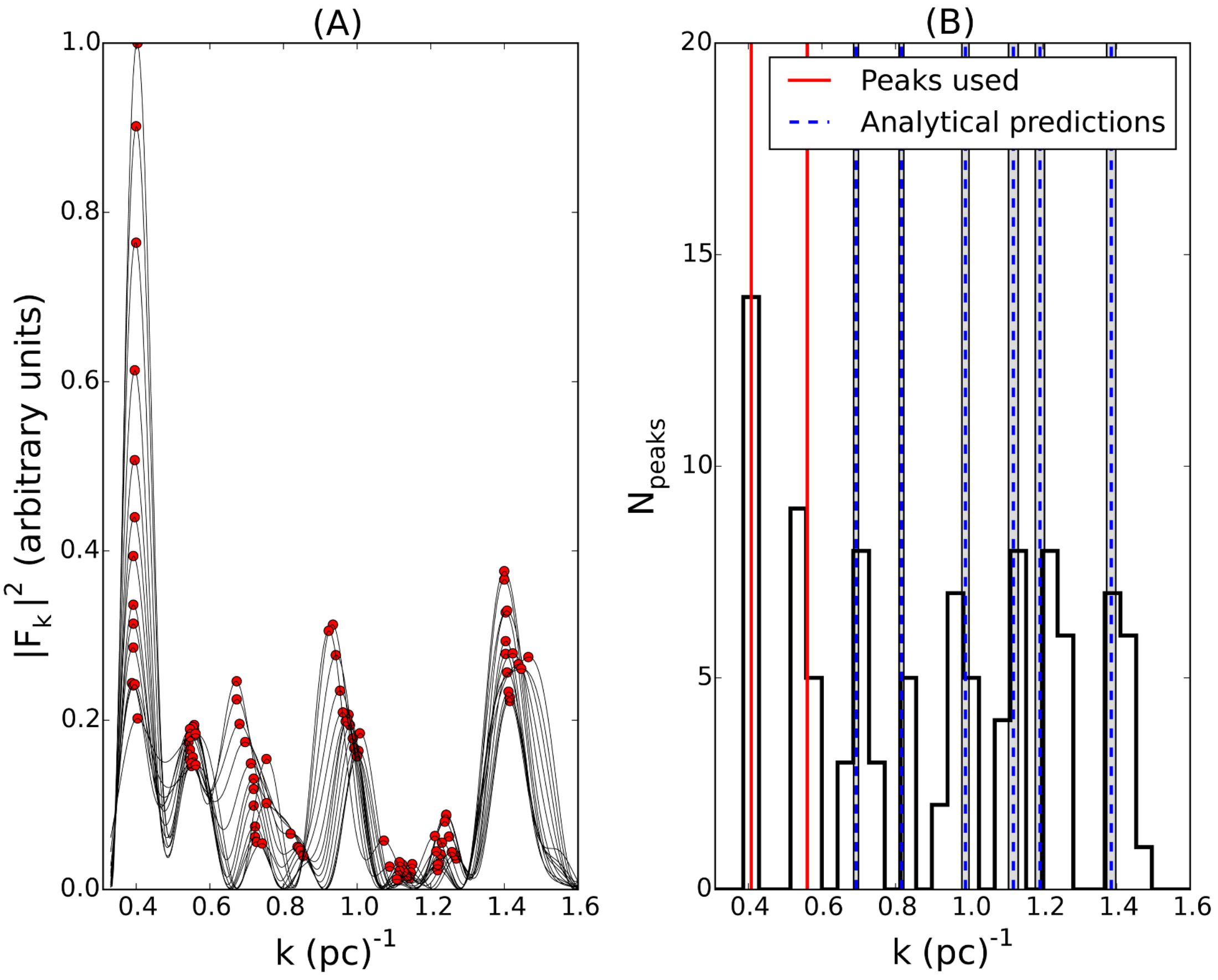

**Fig. S2.**
Normal modes in our simulated cloud. Panels (A) and (B) are the same as Fig. 2 but from our numerical simulation of the 3D model of the cloud.

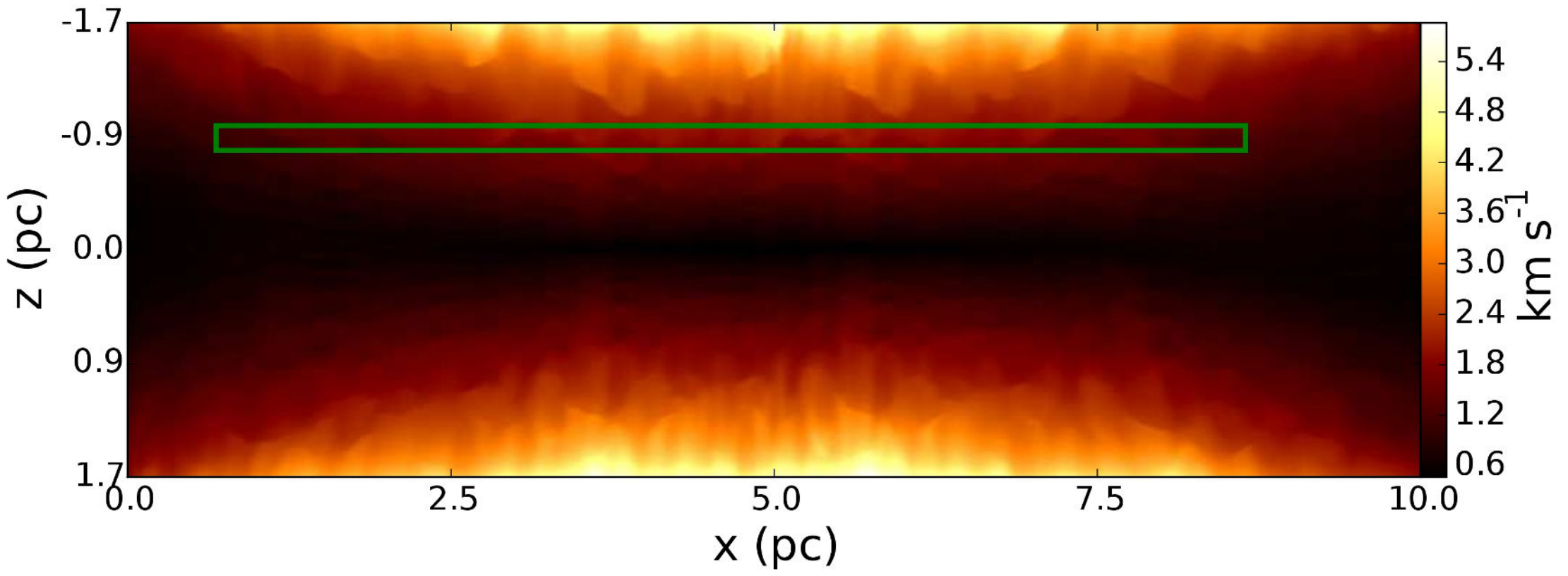

**Fig. S3.**

Boundaries formed in the simulated cloud. Mid-plane map (along the *y* direction) of the propagation speed of fast magnetosonic waves. The propagation speed is clearly higher above and below the main body of the cloud compared to the edges. The green rectangle marks the region within which we take cuts perpendicular to the striations to perform our analysis.

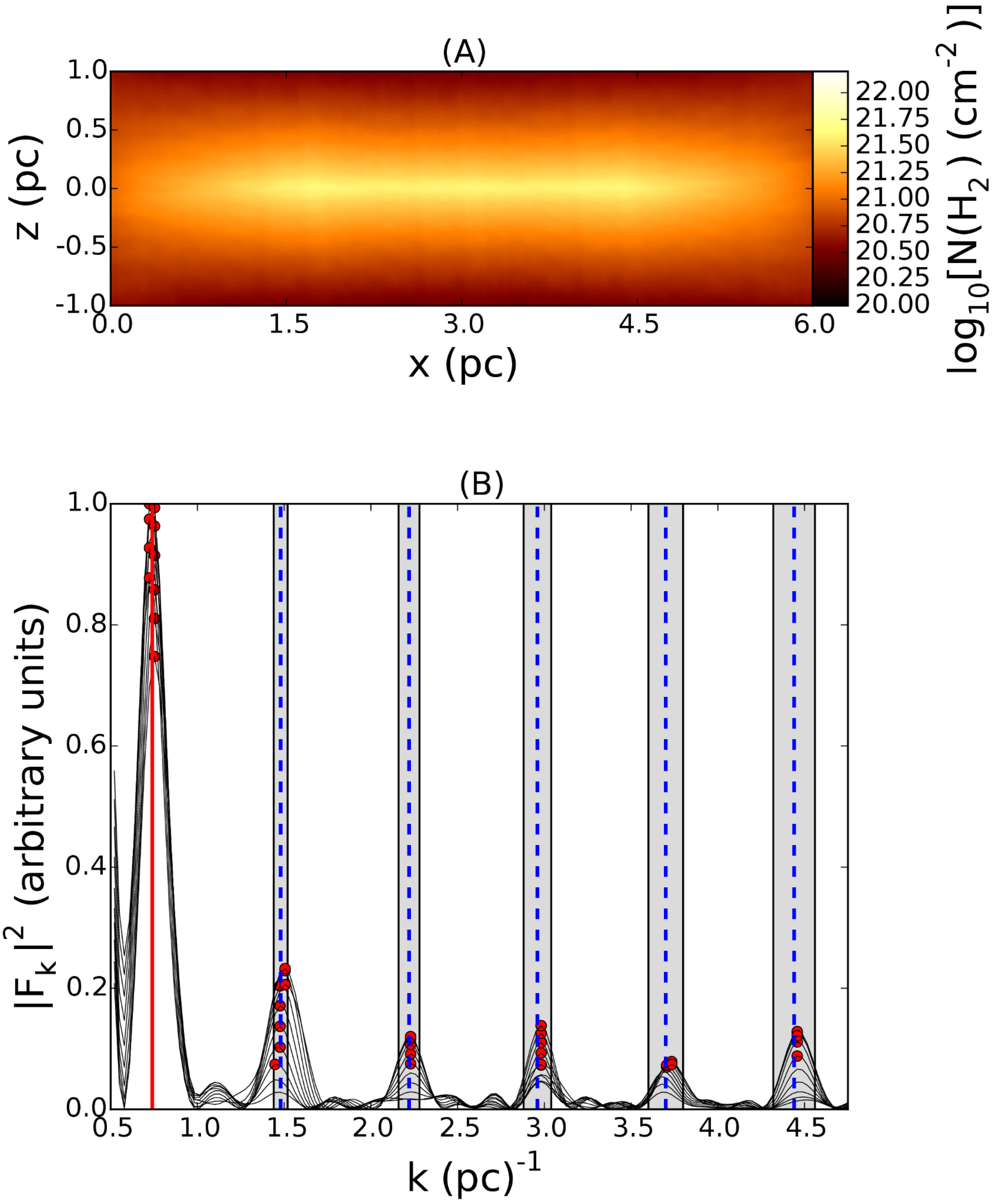

**Fig. S4.**

Normal modes in a cylindrical simulated cloud. Upper panel: column density map from our MHD simulations considering a filamentary/ribbon-like morphology. Lower panel: Power spectra of the cuts perpendicular to the striations considering this morphology. Lines and symbols are the same as in Fig. 2.